\documentstyle[12pt,aaspp]{article}

\begin{document}
\def\siges{\sigma_{es}}
\def\chir{\chi_{R}}
\def\chies{\chi_{es}}
\def\chinu{\chi_{\nu}}
\def\Bnu{B_{\nu}}
\def\taunu{\tau_{\nu}}
\def\tauqnu{\tau_{q\nu}}
\def\Jnu{J_{\nu}}
\def\tauqr{\tau_{q}^{R}}
\def\r*{r_{*}}
\def\n*{n_{*}}
\def\hfkt{h\nu/kT}
\def\hfkT{h\nu/kT}
\def\nesiges{n_{e} \siges}
\def\sigr{\sigma_{R}}
\def\knu{\kappa_{\nu}}
\def\ksnu{\knu^{*}}
\def\drf{\Delta R_{14}}
\def\nel{n_{11}}
\def\The{\theta_e}
\def\apo{\alpha+1}
\def\apt{\alpha+2}
\def\ame{\alpha-8}
\def\thetamax{\theta_{max}}
\def\thetam{\theta_m}
\def\df{\zeta_\nu}
\def\thpnu{\theta_{p_\nu}}
\def\drpf{\Delta R_{p14}}
\def\npel{n_{p11}}
\def\thp{\theta_p}
\def\thm{\theta_m}
\def\tpnu{$T_p(\nu)$}
\def\rpnu{$r_p(\nu)$}

\typeout{}
\typeout{YOU MAY OBTAIN THE FIGURES VIA ANONYMOUS FTP TO mensch.stanford.edu .}
\typeout{THE APPROPRIATE FILE IS figs.9408040.ps.uu . THE FILES HAVE BEEN}
\typeout{processed with the uufiles command, and are about 3MB in size.}
\typeout{}
{\parskip=0cm minus 0.3cm
\title{ Semi-analytic Continuum Spectra of Type II Supernovae and the
Expanding Photosphere Method of Distance Determination}
\author{ Marcos J. Montes\altaffilmark{1} and Robert V.
Wagoner\altaffilmark{2}}}
\affil{Department of Physics and  Center for Space Science and Astrophysics, \\
Stanford University, Stanford, CA 94305-4060}

\altaffiltext{1}{Internet: marcos@mensch.stanford.edu}
\altaffiltext{2}{Internet: wagoner@leland.stanford.edu
\\[0.5cm] SU-ITP--94-26; August,1994; submitted to The Astrophysical Journal}

\begin{abstract}
   We extend the approximate radiative transfer analysis of Hershkowitz,
Linder, and Wagoner (1986) to a more general class of supernova model
atmospheres, using a simple fit to the effective continuum opacity produced
by lines (Wagoner, Perez, and Vasu 1991). At the low densities considered, the
populations
of the excited states of hydrogen are governed mainly by photoionization and
recombination, and
scattering dominates absorptive opacity. We match the asymptotic expressions
for the spectral energy density $\Jnu$ at the photosphere, whose location at
each frequency is determined by a first-order calculation of the deviation of
$\Jnu$ from the Planck function $\Bnu$.
\parskip 0cm
The emergent spectral luminosity
then assumes the form $ L_\nu = 4\pi^2 \r*^2 \zeta^2 \Bnu(T_p) $,
where $T_p(\nu)$ is the photospheric temperature, $\zeta$ is the dilution
factor, and $\r*$ is a fiducial radius [ultimately taken to be the
photospheric radius $r_p (\nu)$]. The atmosphere is characterized by an
 effective temperature $T_e$
($\propto L^{1/4} r_*^{-1/2}$) and hydrogen density $n_H=n_* (r_*/r)^\alpha$ ;
and less strongly by the heavy element abundance and velocity gradient.
 Our major result is the dependence of $\zeta$  on
frequency $\nu$ and the parameters $T_p$, $r_p$, and $\alpha$.

   The resulting understanding of the dependence of the spectral luminosity
on observable parameters which characterize the relevant physical
conditions will be
of particular use in assessing the reliability of the expanding photosphere
method of distance determination. This is particularly important at
cosmological distances, where no information about the progenitor star will
be available. This technique can also be applied to other low-density
photospheres.

\end{abstract}

\newpage

\section{Introduction}

The goal  of this investigation is to gain a broader understanding of the
dependence of the continuum spectral luminosity of Type II supernovae on the
physical conditions near their photospheres as reflected by parameters that are
 in principle observable. We extend the semi-analytic
approach of Hershkowitz, Linder, and Wagoner (1986b, hereafter HLW)
to a more general class of Type II supernova model atmospheres, using a simple
fit to the effective continuum opacity produced by lines (Wagoner, Perez, and
Vasu 1991; hereafter WPV). Understanding the
continuum spectrum allows us to calculate the dilution
factor, a frequency dependent quantity that appears as a correction to the
luminosity due to
the non-blackbody, reduced flux of the continuum spectrum emitted by
supernovae. This dilution factor is arguably the most critical element in the
expanding photosphere method (EPM) of distance determination
(Kirshner \& Kwan 1974; Wagoner 1981; Eastman \& Kirshner 1989; Schmidt,
Kirshner, \& Eastman 1992, hereafter SKE; Wagoner \& Montes 1993;
Eastman, Schmidt, and Kirshner 1994; Schmidt et al. 1994).

The extensions of this work from that of HLW include:
1) spherically-symmetric power law atmospheres; 2) inclusion the UV
opacity of the heavy elements via a simple fit from WPV; 3) allowance for
atmospheric regions of two different types: mostly ionized ($ n_e \cong n_H $;
case I), and mostly neutral ($ n_e \lesssim 0.5n_H $; case N).
In addition, at the low temperatures studied ($5,000 \lesssim T_p \lesssim
20,000$K) it is often found that the Lyman
continuum is in radiative detailed balance. Then collisional
coupling of the $n=1$ and $n=2$ levels of hydrogen becomes the most important
channel determining the departure coefficient of the ground state.
Thus we retain
collisional coupling between these levels, although we have found that it is
otherwise negligible at the low densities of these photospheres ($n_H\lesssim
10^{12}\mbox{ cm}^{-3}$).

We would like to stress that although our results are approximate, they require
no knowledge of the progenitor star and are expressed in terms of observable
parameters which fully characterize the relevant physical conditions near the
photosphere. The analytical aspect of our results also reflects the pedagogical
goal of our investigation.

\newpage

\section{Radiative Transfer Problem}

The assumptions which define our radiative transfer problem are listed below.

1. The velocity gradient is small enough to yield a quasi-static photosphere.
This is a good approximation for most supernovae after a few days.

2. The supernova is spherically symmetric, with the total hydrogen density
given by $n_H = n_*(\r*/r)^{\alpha}$ near the photosphere. Here $n_*$ and $r_*$
are fiducial values.

3. We consider hydrogen dominated systems but include heavier elements
via an effective continuum scattering opacity (from WPV). Hydrogen
photoionization and inverse bremsstrahlung account for the absorptive opacity.

4. We consider only low density photospheres, such that collisions may usually
be neglected (except for the ground state of hydrogen) and for the optical
and UV photons the atmosphere is scattering dominated (by electrons and
mostly iron-peak lines).

5. Radiative detailed balance holds for the hydrogen lines, as verified
by Hershkowitz and Wagoner (1987).

6. We shall adopt the value of 1/3  for the Eddington factor near (and
below) the photosphere since in scattering dominated photospheres the
continuum is formed at an optical depth large enough to make the radiation
field essentially isotropic.

7. Energy is solely transported by radiation.

8. In keeping with our approximate analysis, the hydrogen Gaunt factors are
set to unity.

Assumptions 2,3 \& 4
are more general than those made in HLW,
who also only considered  the case of complete
ionization and did not examine the dilution factor in any detail.
The total opacity and optical depth are
\begin{equation}
\chinu=\chi_{sc} + \knu ,\;\;\;\; \taunu = \int^{\infty}_{r} \chinu (\nu,
r^\prime)dr^\prime \ ,
\end{equation}
where the scattering opacity $\chi_{sc}=\chies +\chi_{lines}$. The
absorptive opacity is due to hydrogen photoionization and inverse
bremsstrahlung:
\begin{equation}
\knu=\sum_{i}(n_{i}-n_{i}^{*}e^{-\hfkt})
\alpha_{i\kappa}(\nu)+n_{e}^{2}\alpha_{ff}(\nu,T)(1-e^{-\hfkt}) \ .
\end{equation}
The LTE populations are given by
$
n_{i}^{*}=n_{e}^{2}\Phi_{i}(T) ,
$ where
$\Phi_i(T) $ is the Saha-Boltzmann function for a pure hydrogen gas.
We assume (and verify) that most of the free electrons come from hydrogen,
giving
\begin{equation}
n_H \cong n_e + (1+d_1)n_{1}^{*}\ ,
\end{equation}
where $d_1$ is the ground-state departure coefficient.

For our extended atmosphere, we use the  Mihalas (1978) formulation of the
transfer equations:
\begin{eqnarray}
\label{eq:mih1}
\partial(f_\nu q_\nu r^2 \Jnu)/\partial \tau_{q\nu} &=& L_\nu/(4\pi)^2 \ , \\
\label{eq:mih2}
\frac{\partial L_\nu}{\partial \taunu} &=& \frac{(4\pi r)^2}{\chinu}
(\knu \Jnu - \ksnu \Bnu ) \ .
\end{eqnarray}
In these equations $f_\nu$ is the Eddington factor,
$d\tauqnu=q_\nu d\taunu ,$ is the `spherical optical
depth', and
$\Bnu = (2h\nu^3/c^2)[\exp(\hfkt)-1]^{-1} $ is the Planck function.
The sphericality factor $q_\nu$ is given by $\ln( r^2 q_\nu ) = \int^{r}_{\r*}
[(3f_\nu -1 )/(r^\prime f_\nu )]dr^\prime + \ln \r*^{2}.$
Since for scattering dominance the total optical depth at the photosphere
$\tau_p(\nu)>1$, we may use $f_\nu (\tau_\nu \gtrsim \tau_p)\cong 1/3$,
giving $q_\nu = (\r*/r)^2$.
Equation (\ref{eq:mih1}) is then simplified to:
\begin{equation}
\label{eq:mih1ap}
\frac{\partial \Jnu}{\partial\tau_{q\nu}}= \frac{3L_\nu}{(4\pi r_*)^2} .
\label{eq:simp}
\end{equation}

The constraint of radiative equilibrium imposes the equivalent conditions
\begin{eqnarray}
\int_{0}^{\infty} L_\nu d\nu \equiv 4 \pi r_{*}^2 \sigma_R T_{e}^{4} &=&
\mbox{constant}\ ,\\
\int_{0}^{\infty}(\knu \Jnu - \ksnu \Bnu )d\nu &=& 0 \ .
\end{eqnarray}
The effective temperature $T_e$ and fiducial radius $r_*$ are parameters
characterizing our model photospheres.
The second constraint is that the atomic level populations are maintained in
statistical equilibrium (not LTE). Because of our assumption of relatively
small velocity gradient broadening, we can take the radiative transitions
between the
bound states to be in detailed balance (Hershkowitz \& Wagoner 1987).
That is, there is a large optical depth in the lines at the continuum
photosphere.
Then the non-LTE
departure coefficients $d_l = (n_l - n_{l}^{*})/n_{l}^{*}$ are given by
\begin{equation}
\label{eq:depco}
d_l \left[ 4\pi\int_{\nu_l}^{\infty} (\alpha_{l\kappa}\Jnu/h\nu)d\nu +
\sum_{j\neq l}^{\kappa}C_{lj} \right]-\sum_{j\neq l}^{L}d_j C_{lj}=
4\pi\int_{\nu_l}^{\infty}(\Bnu-\Jnu)[1-e^{-(\hfkt)}]
(\alpha_{l\kappa}/h\nu)d\nu\ ,
\end{equation}
where $n_lC_{lj}$ is the rate of transitions from level $l\rightarrow j$
induced
by electron collisions (Mihalas 1978).[Here $\kappa$ denotes unbound states,
$\nu_l$ are the threshold ionization frequencies, and $L$ (usually taken
to be 10)is the principal quantum number of the highest level included.]

The appropriate second order equation is obtained by substituting equation
(\ref{eq:mih1ap}) into equation (\ref{eq:mih2}), which yields
\begin{equation}
\label{eq:2ndsimp}
\left(\frac{\chi_\nu}{n_e \sigma_{es}} \right)\left(\frac{r_*}{r} \right)^4
\frac{\partial^2 \Jnu}{\partial \tau_{q\nu}^{2}} = \frac{3}{n_e \sigma_{es}}
(\knu \Jnu - \ksnu \Bnu )\ .
\end{equation}
We rewrite the bound-free and free-free absorption coefficients (setting the
Gaunt factors to unity) as
\begin{eqnarray}
\alpha_{ff}&=&A_{ff}(T)\nu^{-3} \ ,\\
\Phi_l \alpha_{l\kappa}&=&A_l(T)\nu^{-3} \  (\nu>\nu_l) \ .
\end{eqnarray}
The constraint (\ref{eq:depco}) of statistical equilibrium can be incorporated
directly into the radiative transfer equation (\ref{eq:2ndsimp}), yielding for
$\nu_n < \nu <\nu_{n-1}$ our working transfer equation
\begin{eqnarray}
\label{eq:baseeqn}
\left( \frac{r^*}{r} \right)^4 \left( \frac{\chinu}{n_e \siges}\right)
\frac {\partial^2 \Jnu}{\partial \tau_{q\nu}^{2}}  & = &
\frac{3n_e}{\nu^3\siges}
\left\{ \left[ A_{ff}(T)+\sum_{l=n}^{L}A_l(T)\right]
(1-e^{-\hfkt}) (\Jnu-\Bnu) \right. \nonumber \\
 &  &  \left. -\Jnu \sum_{l=n}^{L}A_l(T) \sum_{j=1}^{L}M_{lj}^{-1}
\left[\frac{\int_{\nu_j}^{\infty} (\Jnu - \Bnu) (1-e^{-\hfkt}) \nu^{-4}
d\nu}{ \int_{\nu_j}^{\infty} \Jnu \nu^{-4} d\nu} \right]\right\} \ .
\end{eqnarray}
Here $M_{lj}^{-1}$ is the inverse of the matrix
\begin{equation}
M_{lj}=\left( 1+\sum_{i\neq l}^{\kappa} C_{li}^{\prime}\right) \delta_{lj}
-C_{lj}^{\prime} \ ,
\end{equation}
which is constructed from the relative transition rates
\begin{equation}
C_{lj}^{\prime}=C_{lj}\left[4\pi\int_{\nu_l}^{\infty}(\alpha_{l\kappa}
\Jnu/h\nu)d\nu\right]^{-1} \ .
\end{equation}
\newpage

\section{Approximate Analysis}
We proceed by first matching asymptotic solutions to the transfer
equation at the photosphere. This gives an expression for the emergent
luminosity which depends upon the location of the photosphere. The location is
then determined by solving equation (\ref{eq:baseeqn}) to first order in
$|\Jnu-\Bnu|/\Bnu$. Then the relevant physical quantities at the photosphere
can be related.

For $\taunu \gg \tau_p(\nu) $ we recover LTE populations and a thermalized
radiation field,  giving $\Jnu = \Bnu .$
Integrating equation (\ref{eq:simp}) over frequency, we obtain the familiar
result
\begin{equation}
\label{eq:ttauion}
T^4 = \frac{3L}{16\pi \sigr \r*^{2}} \left(\tauqr +C\right)
    \equiv \frac{3}{4} T_{e}^{4}(\tauqr+C) \ ,\ (C= \mbox{const})\ .
\end{equation}
Here $\tauqr$ is the Rosseland mean spherical optical depth
$(d\tauqr = -q \chir dr)$.
In the other asymptotic limit $\taunu \ll \tau_p(\nu) $,
the spectral luminosity is no longer changing ($L_\nu = L_{\nu}^{o}$),
although scattering will still be important where $\taunu \gtrsim 1$.
Integrating equation (\ref{eq:mih1ap}) in this regime yields
\begin{equation}
\Jnu = 3L_{\nu}^{o} (4\pi \r* )^{-2} (\tauqnu+c_\nu) \ ,\
(c_\nu=\mbox{const.})\ .
\end{equation}
If we now match the asymptotic expressions for $\Jnu$ at $\taunu =
\tau_p(\nu)$ we obtain the emergent luminosity
\begin{equation}
\label{eq:emlum}
L_{\nu}^{o}  =  4\pi \r*^{2} \zeta^2 \pi \Bnu (T_p) \ ,
\end{equation}
where the dilution factor
$\zeta$ is given by $\zeta^2 = (4/3)(\tauqnu+c_\nu)^{-1}_{p}.$ Near the
photosphere we take $\tauqnu\cong (\chinu/\chir) \tauqr
\gg c_\nu \sim C \sim 1, $ giving
\begin{equation}
\label{eq:dfdef}
\zeta  \cong
\left(\frac{\chir}{\chinu}\right)_{p}^{\case{1}{2}}
\left(\frac{T_e}{T_p}\right)^2 .
\end{equation}
The subscript $p$ indicates that the quantity is evaluated at the (frequency
dependent) photosphere.

In order to proceed, we must approximate the dominant opacities in various
 portions of the spectrum. Employing the results of WPV, we adopt the form
\begin{equation}
\label{eq:opacrat}
{\frac{\chinu}{\chies}}=\left\{ \begin{array}{ll}
  1+n_0\Phi_1 \alpha_{1\kappa}/\siges              & \mbox{if $\nu>\nu_{Lyman}$
and $\theta<\theta_{eq}$\ ,} \\
  1+(n_0\Phi_1)^{1/2} \alpha_{1\kappa}/\siges      & \mbox{if $\nu>\nu_{Lyman}$
and $\theta>\theta_{eq}$\ ,} \\
  1+(250\theta^{16}+10\theta^4)(\nu/\nu_{Lyman})^{1/2}   & \mbox{if
$\nu_{Lyman}>\nu> \nu_{inter}$\ ,} \\
  1+(250\theta^{16}+10\theta^4)(\nu/\nu_{inter})^{10} & \mbox{if
$\nu_{inter}>\nu> \nu_{Paschen}$\ ,} \\
  1			      & \mbox{otherwise\ ,}
					  \end{array}
				   \right .
\end{equation}
where $\chies=\nesiges\ ,\theta=5040/T$(K) and $\nu_{inter}=c/3200\AA$.
This gives the results shown in Figure 1.

In the temperature range we are considering, photoionization of  the ground
state of hydrogen usually dominates the total opacity $chinu$ in the
Lyman continuum. With $\taunu (\nu > \nu_{Lyman})\ \gg 1$ at the photosphere
corresponding to lower frequencies, we here assume  an LTE ground state
population. We have used  $n_e\simeq n_{0}$ for case I,
and $n_e \simeq (n_{0}/\Phi_1)^{1/2}$ for case N, matched at $\theta_{eq}.$
We choose $n_{0}$
as a typical density (in the range $n_0 = n_H = 10^{9}\mbox{ cm}^{-3}-
10^{12}\mbox{ cm}^{-3}$). Since this density only enters in the Lyman continuum
it
affects our analysis at longer wavelengths only through the Rosseland opacity,
and as will be shown later this has only a very small effect on the dilution
factor. Since
there is little flux in the Lyman continuum in the temperature range we
are considering, there are only small changes ($\lesssim 5\%$) in
$\chir/\nesiges$ as we vary $n_0$ through the range indicated above.

As a first order approximation to the line scattering opacity due to (mostly
iron peak) heavy elements, we utilize a fit to the results of WPV
[lines 3 and 4 of equation (\ref{eq:opacrat})].
Since the strongest dependence of the scattering opacity is on temperature,
the fit for the Balmer and Paschen continua was produced at the fiducial
values of $n_H = 10^{11}\mbox{ cm}^{-3}, r/v=10$ days, and heavy element mass
fraction $Z=0.02.$ Varying the
coefficients of the fit by a factor of two produced values of $\chir/\chies$
that were at most 15\% greater (doubling the coefficients) or at most
10\% smaller (halving the coefficients). From WPV we see that changing the
coefficients by
this factor would also correspond to varying
$n_H$ from $10^{10}- 10^{13} \mbox{ cm}^{-3},$
$r/v=t-t_0$ from 4 - 30 days, or
$Z$ by a factor of 10. The largest effect these variations will
have on our later analysis is directly through the quantity $\chinu/\chies$
at the frequencies of interest.

In order to assess the importance of the absorptive part of the continuum
opacity due to the lines, we need to
estimate the thermalization parameters and the optical depths $\tau_n$
of the lines. An estimate of the thermalization parameter $\epsilon_n$
(appearing in the source function $S_n = (1-\epsilon_n) \langle J_\nu\rangle_n
+\epsilon_nB_n$ for line $n$ in a two-level atom without continuum) is
provided by equation (5.3) of
WPV, giving $\epsilon_n \lesssim 2\times 10^{-2}$ for
$T \gtrsim 5000\mbox{ K}, n_{e}\lesssim 10^{11}\mbox{ cm}^{-3},$ and
$\lambda \lesssim 6000\mbox{ \AA}$ (where the lines are important).
The ratio of
effective absorptive to total effective continuum opacity in the lines is
then [using equations (3.10) and (3.16) of WPV] $\kappa_{lines}/\chi_{lines}
\sim \langle \epsilon_n(1+\tau_n)\rangle$ if $\langle\epsilon_n\tau_n\rangle
\lesssim 1.$ This is not larger than the ratio of the hydrogen bound-free to
total opacity, using the results of WPV which indicate that $\langle \tau_n
\rangle \sim 1.$ Therefore we shall neglect the contribution of lines
(as well as photoionization of heavy elements) to the absorptive opacity.

The assumption of low-density is implicitly introduced
when we neglect the free-free and bound-free hydrogen opacities
(only in our calculation of $\chinu$ and
$\chi_R$) for frequencies below the Lyman limit. This neglect makes $\chir$
at most about 12\% too small for
$n_H=10^{11}\mbox{ cm}^{-3}.$ This implies a corresponding upper limit on
$n_H$ in order that our assumption of a scattering
dominated system be valid. As one progresses to longer wavelengths,
the ratio of absorption to scattering increases; so our analysis
becomes less accurate in the infrared.

\subsection{Thermal Structure of the Atmosphere}

Following the approach used by HLW, we next need to determine relations
among the variables $T, \ r \mbox{ and } \tau $ for our two ionization regimes
(I,N). A formalism
for the general case may be developed in the following way. The derivative of
equation (\ref{eq:ttauion}) with respect to $r$ gives
\begin{equation}
\label{eq:taur1}
\frac{d \tauqr }{dr}=-\frac{16}{3}\left(\frac{\theta_{e}}{\theta}\right)^{4}
\frac{d\ln \theta}{dr} \ .
\end{equation}
We also have the definition (also only valid at large optical depths)
\begin{equation}
\label{eq:taur2}
d\tauqr/dr = -A (\r*/r)^{2} \nesiges \  ,
\end{equation}
where $A = \chir/\chies \approx A(\theta)$ has a weak dependence on the other
properties of the photosphere: density, velocity, and heavy element abundance.
In what follows, we sometimes take $A(\theta)$ (seen in Figure 1) to be a
slowly varying function, so we may neglect its $\theta$ dependence in
integrations involving more rapidly varying functions over restricted ranges
of temperature.

If we find expressions for
$n_e (\theta ,r)$ we may equate equations (\ref{eq:taur1}) and (\ref{eq:taur2})
 and obtain relations
for $\theta(r/\r*)$ in both the ionized and neutral cases. Implicit in
 the following is the assumption that near (and below) the
photosphere $1+d_1 = n_1/n^{*}_{1} \approx 1$, which we find to be verified by
complete atmosphere calculations.

For the
ionized case  $n_H \cong n_e,$ giving
\begin{equation}
\label{eq:nei}
n_e = n_* (\r*/r)^\alpha \ .
\end{equation}
We then obtain the temperature profile
\begin{equation}
\label{eq:thivsr}
\left(\frac{\theta_e}{\theta}\right)^4 = G_i\left(\frac{\r*}{r}\right)^{\alpha
+ 1} + C_i \ ,
\end{equation}
with $G_i \equiv \threequarters A(\theta) n_* \siges \Delta{r_1}
\mbox{ and }\Delta{r_1}\equiv \r*/(\alpha+1).$ One expects that
$G_i \sim \tau_p,$ so it is reasonable to neglect $C_i \sim 1 $ when
scattering dominance produces a photospheric optical depth  $\tau_p \gg 1.$

For the neutral case $n_H \cong n_{1}^{*}$ gives
$n_e\cong (n_H/\Phi_1)^{1/2},$ so we have
\begin{equation}
\label{eq:nen}
n_e = 2.94\times 10^{10} n_{*}^{1/2} \left(r*/r\right)^{\alpha / 2}
\theta^{-3/4} \exp(-\gamma\theta) \ ,
\end{equation}
where $2\gamma \equiv (h\nu_1/k)/(5040\mbox{ K}) = 31.31.$
For this case we are not able to obtain an exact
analytical expression. However, since we are working in the regime where
$\theta \cong  \theta_e \cong 1 , $ the integral is dominated by the
exponential term and we obtain the approximate form
\begin{equation}
\label{eq:thnvsr}
\left(\frac{\theta_e}{\theta}\right)^4 \theta^{-1/4} \exp(\gamma \theta)
\cong C_n - G_n\left(\frac{\r*}{r}\right)^{1+ \alpha /2}\ ,
\end{equation}
where $G_n\equiv 8.63 \times 10^{10} A(\theta) n_{*}^{1/2} \siges \Delta r_2
 \mbox{ and }\Delta r_2 \equiv 2 \r*/(\alpha + 2).$

Since we expect the function $\theta(r)$ for the ionized and neutral regimes
to match
at $n_e/n_H \cong 0.5,$ we determine the location $\theta_m$ of the match by
solving for the intersection of equation (\ref{eq:thivsr}), with $C_i=0$
and $n_H \Phi_1 =2$:
\begin{equation}
\label{eq:thmeq}
2 \gamma \theta_m + \frac{3-5\alpha}{2\alpha+2} \ln \theta_m  =
48.90- \ln n_* + \frac{\alpha}{\alpha+1}(\ln G_i - 4\ln \theta_e ) \ ,
\end{equation}
with $r_m/\r* = G_{i}^{1/(\alpha+1)} (\theta_m/\theta_e)^{4/(\alpha+1)}.$
Since $\ln \theta_m \cong \ln \theta_e \cong 0 $, we may approximate
equation (\ref{eq:thmeq}) as
\begin{equation}
\label{eq:apthm}
\theta_m \cong 3.194\times 10^{-2}\left[48.90-\ln n_* +
\frac{\alpha}{\alpha+1}\ln G_i \right] \ .
\end{equation}
Equation (\ref{eq:apthm}) is used only to indicate the major dependencies of
$\theta_m,$ and is never used in our numerical work. From equation
(\ref{eq:thmeq}) we find that $\theta_m$ is essentially a universal function
of a particular combination of model
parameters, indicated in  Figure 2.

Also matching equation (\ref{eq:thnvsr}) at this point determines the
integration constant  to be
\begin{eqnarray}
C_n & = & \left(\frac{\theta_e}{\thetam}\right)^4\thetam^{-1/4} \exp(\gamma
\thetam) +
     G_n G_{i}^{-(\alpha+2)/(2\alpha+2)}
      \left(\frac{\theta_e}{\thetam} \right)^{2(\alpha+2)/(\alpha+1)} ,
\label{eq:cneq}
\\
 \mbox{} & = & 9.30 \times 10^4
\left(\frac{\theta_e}{\thetam}\right)^{\frac{2\alpha+4}{\alpha+1}}
\nel^{\frac{-1}{2(\alpha+1)}} (4.99
A(\thetam)\drf)^{\frac{\alpha}{(2\alpha+2)}} \left[ \sqrt{2}
\thetam^{-1}+\frac{\gamma}{2} \frac{\apo}{\apt} \right] .
\label{eq:cneq2}
\end{eqnarray}
We have introduced the dimensionless parameters
$\nel=n_*/10^{11}\mbox{ cm}^{-3}$ and $\drf= \Delta r_1/10^{14} \\\mbox{cm} =
[r_*/)\alpha+1)]/10^{14}\mbox{ cm}.$
A more aesthetically pleasing location of the match is provided by requiring
that both the temperatures and their derivatives match. This approach yields a
value of $f_{m}\equiv (n_e/n_H)_{match}= 0.46 - 0.51$ for a wide range of
models.

In Figure 3 we plot the (inverse) temperature
structure of the atmosphere with the choice $\Delta R_{14}=1.0,\  n_{11}=1,$
 and $\alpha=8$ (for three different values of the effective temperature).
As shown in this figure, these models also have an asymptotic temperature
$\thetamax$ given by equations (\ref{eq:thnvsr}) and (\ref{eq:cneq}) as
\begin{equation}
\exp(\gamma\thetamax) \thetamax^{-17/4} = \exp(\gamma\thetam) \thetam^{-17/4}
\left[ 1+ \frac{\gamma\thetam}{2\sqrt{2}}\left(\frac{\apo}{\apt}\right)\right]
 = C_n \theta_{e}^{4}\ .
\end{equation}
In addition, we obtain a minimum value of the ionization fraction
$(n_e/n_H)_{min} \sim 0.2$ for
these atmospheres with a wide range of model parameters.
This validates our neglect of the contribution of
elements other than hydrogen to the electron density. However, it is
important to remind the
reader that these temperature structures are only valid for
$\taunu \gtrsim 1 .$ At small  optical
depths the temperature and ionization are governed by the radiation field
(formed at large optical depths), which tends to drive the level populations
far from their LTE values.

\subsection{Location of the Photosphere}

Following the approach of HLW, we now operate on our master equation
(\ref{eq:baseeqn}) with $\int_{\nu_k}^{\infty}\nu^{-1}d\nu$, obtaining
\begin{eqnarray}
\int_{\nu_j}^{\infty}\frac{\chinu}{\nesiges}\left(\frac{r_*}{r}\right)^4
\frac{\partial^2 \Jnu}{\partial \tauqnu^2} \nu^{-1} d\nu \nonumber  & = &
\frac{3 n_e}{\siges} A_{ff} \left[ D_j + \sum_{l=1}^{j-1} A_{l}^{\prime}
\left( D_l - E_{l}^{\prime} \sum_{k=1}^{L} M_{lk}^{-1}D_k/E_{k}^{\prime}
\right) \right. \\
 &  & \left. \mbox{} + \sum_{l=j}^{L} A_{l}^{\prime}\left( D_j -
E_{j}^{\prime}\sum_{k=1}^{L} M_{lk}^{-1}D_k/E_{k}^{\prime}\right) \right]
\nonumber\\
 & \equiv & \frac{3 n_e}{\siges}A_{ff} \sum_{k} Q_{jk}D_k ,
\end{eqnarray}
where
\begin{eqnarray}
\nonumber
D_k  &= & \int_{\nu_k}^{\infty} (\Jnu - \Bnu ) (1-e^{-\hfkT})^{-1} \nu^{-4}
d\nu , \\
\nonumber
E^{\prime}_{k} & = & (c^2/2h)\int_{\nu_k}^{\infty} \Jnu \nu^{-4} d\nu ,
\end{eqnarray}
and $A^{\prime}_{l}=A_l/A_{ff}.$ Next, we invert the above
matrix equation to obtain the first-order quantities $D_k$ and substitute
the expression into the equivalent quantities
in equation (\ref{eq:baseeqn}).

We next approximate the resulting transfer equation as the photosphere
is approached from below by replacing all occurrences of
$\Jnu$ with $\Bnu$, but keeping terms in $\Jnu-\Bnu$ as the \newpage
{\noindent first} order correction. This yields (for $\nu_n<\nu<\nu_{n-1}$)
\begin{eqnarray}
\label{eq:master}
\lefteqn{\left(1+\sum_{l=n}^{L}A_{l}^{\prime}\right)(1-e^{-\hfkT})(\Jnu-\Bnu)
=}\nonumber\\
& & \frac{\nesiges}{3 A_{ff}} \left[ \nu^3 N(\nu,\theta) +\Bnu
\sum_{l=n}^{L}A_{l}^{\prime} \sum_{j=1}^{L} M_{lj}^{-1} \frac{\sum_{k=1}^{L}
Q_{jk}^{-1}
\int_{\nu_k}^{\infty} N(\nu,\theta)
\frac{d\nu}{\nu} }{E_{j}^{\prime}} \right]\ ,
\end{eqnarray}
with
\begin{equation}
 N(\nu,\theta) = \left(\frac{\chinu}{\nesiges}\right) q^2 \frac{\partial^2
\Bnu}{\partial
\tauqnu^{2}}\ ,
\end{equation}
and with the integral in (dimensionless) $E_{l}^{\prime}$ now over $\Bnu$
instead of $\Jnu$.
Since all the Gaunt factors are unity,
\begin{eqnarray}
\nonumber
A_{ff} & = & 5.20 \times 10^6 \theta^{1/2} cm^{5} s^{-3} \ ,\\
A_{l}^{\prime} & = & 62.5 l^{-3} \theta \exp(31.31\theta/l^2) \ .
\end{eqnarray}

The structure of the matrices $Q_{ij}$ and $M_{ij}$ depends upon the
approximations employed. Naively, in the limit of very low densities
collisions may be neglected, giving
$M_{ij}=\delta_{ij},$ as in HLW. However, since the Lyman continuum is usually
in radiative detailed balance due to
its large absorptive opacity, we keep the collisional coupling between the
$n=1$ and 2 levels of hydrogen. This modifies $M_{ij}^{-1}$ by the inclusion
of the term $u=C_{12}^{\prime}/(1+C_{12}^{\prime})$ in
two elements, $M_{11}^{-1}=1-u \mbox{ and } M_{12}^{-1}=u,$ with the rest of
the matrix being unchanged. Because of the relative lack of photons in the
Lyman continuum at the temperatures of interest, $u$ is not necessarily small,
even at our low densities. We have explored the
range $0\leq u\leq 1$, and the results we report for the optical and IR do not
change as we vary $u.$ The largest change occurs at the highest
temperatures ($T \gtrsim 12,000\mbox{ K}$), but even then the changes are
relatively small and are restricted to the Lyman continuum.

We now investigate the properties of the photosphere in terms of the
major model parameters:
$\theta_e,\  \Delta R_{14},\  n^{*}_{11},\mbox{ and } \alpha .$
In keeping with our first-order analysis, we may use the zeroth order
expression (\ref{eq:ttauion}) for the relation between optical depth and
temperature in simplifying the expression for
$\partial^2 \Bnu /\partial \tauqnu^{2}$ .
We obtain
\begin{equation}
\partial^2 \Bnu /\partial \tauqnu^{2}\cong (9/16)(\chir/\chinu)^2
(\theta/\theta_e)^8 \Bnu f(\hfkT) \ ,
\end{equation}
where
\begin{equation}
f(x)=\frac{[x-5+(x+5)e^{-x}]x}{16(1-e^{-x})^2} \ .
\end{equation}

The first order deviation of the average intensity from blackbody is then
given by
\begin{equation}
\label{eq:masters}
\frac{|\Jnu-\Bnu|}{\Bnu}= \left[\frac{2.778\times
10^{10}}{n_{e}(\mbox{cm}^3)}\right] \left(\frac{\r*}{r}\right)^4
\frac{\theta^{9/2}}{\theta_{e}^{8}} S_2(\theta) S_3(\theta,\nu)
\end{equation}
in the interval  $\nu_n < \nu < \nu_{n-1} $, with
\begin{eqnarray}
\label{eq:sdefs}
S_1(\theta,\nu) & = & \chir/\chinu\ , \nonumber \\
S_2 (\theta) & = & \chir /\chies\ ,\\
S_3 (\theta,\nu) & = & \frac{|y^3 f(y) S_1(\nu,\theta) + Q_n|
}{(1-e^{-y})\left( 1+ \sum_{l=n}^{L}A_{l}^{\prime} \right)},\nonumber \\
Q_n & \equiv & \sum_{l=n}^{L} A_{l}^{\prime} \sum_{j=1}^{L}
(M_{lj}^{-1}/E_{j}^{\prime}) \sum_{k=1}^{L} Q_{jk}^{-1} \int_{y_k}^{\infty}
S_1(\nu,\theta) f(x)(e^x-1)^{-1}x^{2}dx \ ,
\end{eqnarray}
and $y\equiv h\nu/kT.$
In order to further utilize these equations, we must determine the
function $n_{e}^{-1} (r^*/r)^4,$ which is straightforward in the two limits
of ionization.

For the ionized case (I) we may write equation (\ref{eq:masters}) in the
following form
after using equations (\ref{eq:nei}) and (\ref{eq:thivsr}) to obtain
$n_e(\theta)$
and $r(\theta):$
\begin{equation}
|\Jnu - \Bnu|/\Bnu  =  F_1(\drf,\nel,\alpha) F_2(\The,\alpha)
F_3(\theta,\nu,\alpha) \ ,
\label{eq:sep}
\end{equation}
where
\begin{eqnarray}
F_1 & = & (4.99 n_{11})^{-5/(\alpha + 1)} \Delta R_{14}^{(\alpha - 4)/(\alpha
+1)}\ , \nonumber \\
F_2 & = & \theta_{e}^{(8-12\alpha)/(\alpha+1)}\ , \\
F_3 & = & 1.386 S_{2}^{(2\alpha -3 )/(\alpha+1)} \
\theta^{(17\alpha-23)/(2\alpha+2)} S_3(\nu, \theta) \ . \nonumber
\end{eqnarray}
In the case where $M_{lj}^{-1}= \delta_{lj}$ (no collisions),
$\alpha \rightarrow \infty $ (sharp atmosphere), and
$\chinu = \nesiges$ (no line scattering), this case corresponds
to the treatment of HLW.

We estimate  the location
of the photosphere at each frequency as the depth at which the mean
intensity differs from the
Planck function by of order the Planck function,
$ |\Jnu - \Bnu | \approx \Bnu $. As in HLW, we fit a smooth function
through the temperature region where $\Jnu-\Bnu$ changes sign.
This procedure should reflect the smooth dependence of the location
of the photosphere on our model parameters. Numerical
calculation of  $S_1,\  S_2,\mbox{ and } S_3$ allows us to determine the
location of the photosphere, represented by $\theta_p$
(for fixed $\nu$ and $\alpha$), as the particular function of model parameters
$\The,\ \drf,\mbox{ and }\nel$ indicated in the above functions $F_1$ and $F_2$
and shown in Figure 4a.

The neutral case (N) is much more difficult. Inspection of equations
(\ref{eq:thnvsr}) and (\ref{eq:cneq2}) shows us that we cannot separate the
dependencies as completely  as in equation (\ref{eq:sep}) since model
parameters occur in both
$C_n \mbox{ and } G_n$. Nevertheless, using equations (\ref{eq:nen}) and
(\ref{eq:thnvsr}), equation (\ref{eq:masters}) now assumes the form
\begin{eqnarray}
\nonumber
\label{eq:neutmast}
\lefteqn{\frac{\Jnu-\Bnu}{\Bnu}  =  10.85 \left( \drf \frac{\apo}{\apt}
\right)^{\frac{\ame}{\apt}} \left( 3.63\times10^6 \nel^{\frac{1}{2}}
\right)^{\frac{-10}{\apt}}  \The^{-8}}\hspace{1.0in} \\
 & & \times  S_{2}^{\frac{2\alpha-6}{\apt}}\theta^{\frac{21}{4}}e^{\gamma
\theta} \left( C_n - \The^4 \theta^{\frac{-17}{4}}e^{\gamma
\theta}\right)^{\frac{8-\alpha}{\apt}} S_3 \ .
\end{eqnarray}

Figure 4b is a plot of the numerically calculated
$\theta_p$ [from setting equation (\ref{eq:neutmast}) equal to unity] for a
range of model
parameters, with the constraint that $\thetam \leq \theta_p \leq \thetamax.$
If $T_p$ is only slightly less than $T_m$, then $\theta_p$ should predominantly
be a function of the combination
$X=\The^{(10\alpha-8)/(\alpha+1)} \drf^{(8-\alpha)/(2\alpha+2)}
\nel^{9/(2\alpha+2)}.$
Notice that except for the special case of $\alpha=8$, there is some
spread in the relation. However, for $\alpha=6$ we obtain a narrow spread in
the relation $\theta_p \mbox{ vs. } X,$ corresponding to
$\theta_p/\thetam < 1.1.$ The spread is much greater for $\alpha=10.$

\subsection{Calculation of the Dilution Factor}

Now we have all the ingredients necessary to calculate the dilution
factor $\zeta$ from equation (\ref{eq:dfdef}) in terms of model parameters.
In the previous section we indicated how the function
$\theta_p (\theta_e, \Delta R_{14}, n_{11}, \alpha; \nu )$
is obtained. Thus  for any desired set of frequencies we may replace
 $\theta_e$ through this function in equation (\ref{eq:dfdef}), which yields
$\zeta = \zeta(\theta_p,n_{11}, \Delta R_{14}, \alpha,\nu)$.

Thus far, the model parameters have referred to a fiducial radius $r_*,$
and thus are not observed quantities. We now calculate the dilution factor
in terms of parameters evaluated at the photosphere, which are (potentially)
observable. In addition to $\theta_p(\nu)$ we employ the
photospheric radius $ r_p (\nu)= v_p(t-t_0),$ where the velocity $v_p(\nu)$
at the photosphere is obtained from analysis of an appropriate
line profile (most reliably from the sharp minimum of a
weak line). The photospheric density is then $n_H(r_p)=n_p(\nu).$

We now choose our fiducial radius to be the photospheric radius, so that
$r_*=r_p(\nu)=r_p14 \times 10^{14}$ cm and $n_*=n_p(\nu)=n_{p11}\times 10^{11}
\ \mbox{cm}^{-3}.$ We also define the photospheric scale height
$\drpf = r_{p14}/(\apo).$ Setting equation (\ref{eq:masters}) equal to
unity, we then obtain relations among the photospheric quantities:
\begin{equation}
\label{eq:nptpi}
0.2779 n_{p11}^{-1} \The^{-8} \theta_{p}^{9/2} S_2(\theta_p) S_3(\theta_p) =1
\end{equation}
for the ionized case and
\begin{equation}
\label{eq:nptpn}
2.988\times 10^{-6} n_{p11}^{-1/2} \The^{-8} \theta_{p}^{21/4}
e^{\gamma\theta_p} S_2(\theta_p) S_3 (\theta_p,\nu) = 1
\end{equation}
for the neutral case. We use these two equations to eliminate $\npel$ from
equations (\ref{eq:thivsr}) and (\ref{eq:thnvsr}) at $r=r_*=r_p.$

We also use equation (\ref{eq:dfdef}) to eliminate $\The$ in favor of $\zeta$,
yielding expressions for the dilution factor for both cases in terms of
photospheric quantities. For the ionized case, we obtain
\begin{equation}
\label{eq:dfi}
\zeta =\zeta_i = 0.947\drpf^{-1/6} \theta_{p}^{7/12} S_{1p}^{1/2} S_{2p}^{-1/3}
S_{3p}^{-1/6} \ ,
\end{equation}
and for the neutral case we obtain, for $\zeta=\zeta_n,$
\begin{eqnarray}
\label{eq:dfn}
\nonumber
\lefteqn{\zeta^{-6} \drpf^{-1} \theta_{p}^{5/2} S_{1p}^{3} S_{2p}^{-1}
S_{3p}^{-1}
+ 10.85 \left(\frac{\apo}{\apt}\right) S_{2p} = S_{2m}\left[10.85
\left(\frac{\apo}{\apt}\right)
+\frac{1.96}{\thetam}\right] }\\
 & & \times
\left[1.498\times 10^{5} \thetam^{-2} S_{2m}^{-1/2}
\zeta^{-5} \drf^{-1/2} \theta_{p}^{19/4} e^{-\gamma \theta_p} S_{1p}^{5/2}
S_{2p}^{-1} S_{3p}^{-1}\right]^{\frac{\apt}{\apo}},
\end{eqnarray}
where the subscripts $p \mbox{ and } m$ indicate that the function is evaluated
at $\theta_p \mbox{ and } \thetam,$ respectively. These equations represent
the most important results of this paper.
The dilution factors for the mostly ionized and mostly neutral cases are
presented
in Figures 5a and 5b.

For the ionized case we have a universal dependence on photospheric scale
height, with a unique dependence on photospheric temperature for each
wavelength, as shown in Figure 5a. The decrease in the dilution factor
(at fixed scale height) with decreasing temperature and wavelength is due to
the increase in UV opacity as the temperature drops, although the
(Rosseland) mean opacity is barely changing.

For the neutral case there is no such simple scaling. In order to obtain the
dilution factor from equation (\ref{eq:dfn}) we employ the same steps
as above to eliminate
$n_{*},\ \Delta r_1 , \mbox{ and }\The$ in equation (\ref{eq:thmeq})
for $\thetam$ in favor of
$\zeta,\ \drpf, \mbox{ and }\theta_p$. We then iteratively solve equation
(\ref{eq:dfn}), which is easily done because $\thetam$ is a weak function of
the parameters, as indicated in Figure 2. In addition, we only
accept solutions for which  $\thetam \le \theta_p \le \theta_{max}$.
For $\alpha>4$
it is found that there are either two or no solutions for $\zeta_n.$
The smaller root always occurs where
$\theta_p/\theta_{max} \gtrsim 0.99,$ which corresponds to  the outer
crossing of the $n_e/n_H=0.5 $ and temperature curves, as seen in
Figure 3. We see that the physical location of this root is in the outer
region of the atmosphere that is not accurately modelled.
In addition, this smaller root does not match $\zeta_i$ as the photosphere
becomes ionized.
For these reasons we always choose the larger of the two values of $\zeta_n.$

Figure 5b shows how
$\zeta_n$ varies with photospheric temperature, scale height, and wavelength;
and its match to $\zeta_i.$ We also find that for $\thp$ only slightly
greater than $\thm$ we obtain a dependence $\zeta_n \propto \drpf^{-1/10},$
as expected when the first term in equation (\ref{eq:dfn}) is negligible.
For fixed scale height $\drpf = r_{p14}/(\apo),$ the behavior of the dilution
factor shown in Figure 5b is relatively insensitive to the value of
$\alpha$ (at least for the range $6\leq\alpha\leq10$ investigated).

A particular frequency dependence that we have investigated is that across
the Balmer ionization threshold at $\lambda=3646\mbox{ \AA},$ motivated by the
dependence on $\Delta R$ found by Hershkowitz, Linder, \& Wagoner (1986a).
We find that the absolute value of the fractional jump in the
photospheric temperature (corresponding to the results in Figures 4a and 4b)
is less than 0.03 for our ranges of the photospheric parameters. The
fractional jump in the dilution factor (corresponding to the results in
Figures 5a and 5b) is likewise found to be less than 0.04. Therefore it
appears that the line scattering (and atmospheric extension) has reduced the
jump from the values found with only electron scattering in a sharp
photosphere by Hershkowitz, Linder, \& Wagoner (1986a).
\newpage

\section{Discussion}

The determination of the distance of a supernova via EPM would proceed as
follows within the above formulation, which assumes nothing (except
spherical symmetry) about the nature of its progenitor. (The other assumptions
we have listed in section 2 can be checked after one obtains the photospheric
conditions from our analysis.) From each observed spectrum, the photospheric
temperature \tpnu\ is first estimated by fitting Planck functions to the
continuum in the neighborhood of various frequencies. As has been indicated
above, the radius \rpnu\ of the photosphere is obtained from (the weaker) line
profiles. The remaining parameter to be determined is $\alpha,$ the total
hydrogen density power-law index.

A comparison of detailed model atmospheres with spectra of SN1987A obtained
during days 2-10 after explosion led Eastman \& Kirshner (1989) to conclude
that $7\leq\alpha\leq11.$ It was found that the UV continuum as well as
the line shapes were sensitive to this parameter, although not greatly so
in the range indicated. However, Branch (1980) has shown that the effects of
optical depth (i.e., heavy element abundance) and density profile on
line shapes may be difficult to separate. In addition, nonLTE effects
(which he did not include) will be important for some lines.

Since it may be difficult to determine $\alpha$ accurately  from lines,
let us consider whether it might be obtained in another way from
observations of the continuum. If one follows a fiducial volume element  which
always contains the same nuclei, its radius and density soon obey $r_* \propto
t-t_0$ and $n_* \propto (t-t_0)^{-3}.$ It then follows that the corresponding
photospheric quantities are related by $n_p(\nu,t)\propto(t-t_0)^{(\alpha-3)}
r_p^{-\alpha}(\nu,t).$ From this relation, one then sees that
\begin{equation}
\label{eq:alpeqn}
\alpha(t)=\left[3+\frac{\partial \ln n_p}{\partial \ln (t-t_0)}\right]
\left[1-\frac{\partial \ln r_p}{\partial \ln (t-t_0)}\right]^{-1}\ .
\end{equation}

In principle, one could obtain the function $n_p(\thp,r_p,\alpha)$ by using
equation (\ref{eq:dfdef}) to write equations (\ref{eq:nptpi}) and
(\ref{eq:nptpn}) in the form
\begin{equation}
\label{eq:npzi}
n_p=2.78\times10^{10} \zeta_{i}^{4} \thp^{-7/2} S_{1p}^{-2} S_{2p} S_{3p}
\mbox{ cm}^{-3}
\end{equation}
for the ionized case and
\begin{equation}
n_p=0.893 \zeta_{n}^{8} \thp^{-11/2} e^{2\gamma\thp} S_{1p}^{-4}S_{2p}^{2}
S_{3p}^{2} \mbox{ cm}^{-3}
\end{equation}
for the neutral case. Our results for the dilution factor
$\zeta(\thp,r_p,\alpha;\nu)$ would then be inserted into these relations.
For instance, one would then obtain $n_{11}=0.224 \drpf^{-2/3} \thp^{-7/6}
S_{2p}^{-1/3} S_{3p}^{1/3}$ for the ionized case. We also note that SKE quoted
the same relation between $\zeta$ and $n_p$ as seen in equation
(\ref{eq:npzi}), for fixed $\thp .$ Of course, since $n_p$ depends upon
$\alpha ,$ the solution of equation (\ref{eq:alpeqn}) would require iteration,
using values of $\thp$ and $r_p$ obtained at various epochs. One could assume
that $\alpha(t)$ was  a slowly varying function. However, a test of the
practical viability of this method is beyond the scope of this paper.

Once the dilution factor $\zeta(\thp,r_p,\alpha;\nu)$ has been determined,
its relation to the luminosity [equation (\ref{eq:emlum}), with $r_*=r_p$]
could be applied in two steps. First, because of the frequency dependence
of the dilution factor, the previous estimate of the photospheric temperature
should be improved by iterating the fit of equation (\ref{eq:emlum}) to the
observed continuum. Second, once the fundamental parameters \tpnu, \rpnu, and
$\alpha$ have been determined, equation (\ref{eq:emlum}) can be employed to
obtain the luminosity distance. Of course, if the supernova is in the Hubble
flow, the redshift $Z$ (of the supernova or parent galaxy) will then produce
a value of the Hubble constant [and for redshifts $Z\geq 0.4,$ the deceleration
parameter (Wagoner, 1977)].

Another important question is the sensitivity of the dilution factor
to the coefficients of the line-scattering opacity. We have both increased and
decreased the coefficients by a factor of two. The dilution factor is most
affected in the Balmer continuum (where the effects of the line opacity are
strongest). At the highest temperatures, the dilution factors in the
Paschen continuum only vary by a few percent. However, as the scattering
increase at lower temperatures, we see much larger affects. Equations
(\ref{eq:dfi}) and (\ref{eq:sdefs}) show that $\zeta_i \propto S_{1p}^{1/3},$
so doubling the coefficients leads to variations of about 25\% when the
line scattering is greater than the electron scattering. However, doubling the
coefficients corresponds to increasing the heavy element abundance by about
an order of magnitude, as shown by WPV. In principle, this abundance can be
determined by the behavior of the UV portion of the spectrum.

Like the heavy element abundance, the extinction of the spectrum by our galaxy
and the supernova parent galaxy must be determined before a reliable luminosity
and distance is obtained. If the dependence of the extinction on wavelength
is universal, then its magnitude can be obtained by including this dependence
in the fit of the spectrum. However, this procedure only becomes reliable and
sensitive if the observations extend from the IR to at least the near UV.

A major challenge that faces us is to reconcile our dilution factor
with that obtained by SKE. There is qualitative agreement in the ionized
regime, although the temperature dependence is somewhat different. However,
in the recombination era our dilution factor decreases as the temperature
decreases, whereas the SKE dilution factor [as well as that recently
obtained by Baron, et al. (1994)] increases. We can understand our
result based on the fact that as the ionization fraction decreases, the
opacity due to absorption (proportional to $n_{e}^{2}$) decreases faster
than the electron and line scattering opacity (roughly proportional to
$n_e$), producing a more dilute radiation field. Another point to make in
comparing our work to that of SKE is that the color temperature they
determined may not correspond to our photospheric temperature. These
temperatures
should be similar as long as there is no net flux in the lines, the
frequency dependence of the dilution factor is negligible, and a continuum
can be uniquely determined. The increase in the density of lines toward the
UV make the determination of a photospheric temperature potentially more
difficult in the B and U bands.

Some other crucial questions remain, which can only be answered by
detailed comparisons between the spectral luminosity predicted by this
method and the observed flux of a variety of Type II supernovae at various
epochs. This is the next step in our program. Some of these questions are:

\noindent 1) How valid are the approximations (such as scattering dominance)
that we have made in this analysis? (If the dilution factor approaches unity,
as probably occurs in the infrared wavelengths, the photosphere is no longer
scattering dominated.)

\noindent 2) How closely does our definition of the photosphere correspond to
reality?

\noindent 3) How tightly can the luminosity be determined from the ranges
of the parameters constrained by the spectral fit?

In spite of these uncertainties, we believe we have developed a
potentially useful tool for determining the luminosity of Type II
supernovae directly from observables in a model-independent manner.
While the results for the ionized regime seem fairly robust, more
work is needed to understand more fully the applicability of our results in
the neutral regime.

\acknowledgements

This research was supported by a grant to the Supernova Intensive Study
group by the Space Telescope Science Institute, which is operated by AURA,
Inc., under NASA contract NAS 5-26555. M.J.M. was supported by the NASA
Graduate Researchers Program through grant NGT 70194-52, and by a fellowship
from the ARCS foundation. R.V.W benefitted from discussions at the 1993
Aspen Physics Center summer workshop on supernovae.
\newpage

\begin{figure}
\caption{
 The temperature dependence of the ratio of Rosseland mean to total opacity,
choosing fiducial parameters
$n_H=10^{11} \mbox{ cm}^{-3},\  r/v=10\mbox{ days, and } Z=0.02$.
The curves correspond to the following wavelengths, from top to bottom:
9000 \AA, 7000 \AA, 5500 \AA, 4400 \AA,
3650 \AA, and 3000 \AA. The top curve also gives
 $\chir/\chies$, since for $\lambda > \lambda_{Paschen}, \chinu \approx
\chies$.}

\caption{The (tight) dependence of the matching temperature $\thetam$ on a
particular combination of
model parameters. This plot has points for $\alpha=6,\ 8,\mbox{ and } 10$.
The range of the other parameters is $0.01\leq \drf \leq 100\ ,
\ 0.01\leq n_11 \leq 10,\ 0.6615\leq \The\leq 1.26 .$}

\caption{
The (inverse) temperature profile as a function of the scaled radius
for model atmospheres with $\alpha=8,\  n_{11}=1 \mbox{ and } \drf=1;$ for
several values of effective temperature.
Note that the curves match smoothly at the fitting point $n_e/n_H=0.5.$
The ionization fraction remains greater than 10\% for a wide range of models.
temperatures.}

\figurenum{4a}
\caption{
The dependence of the photospheric temperature, in the ionized case,
on a combination of model parameters for three choices of wavelength.
The dashed and
solid curves are for $\alpha=10 \mbox{ and } \alpha=6$, respectively. The
results for $\lambda= 3650\mbox{ \AA} \mbox{ and }
5500\mbox{ \AA}$ are  essentially identical to that for 4400 \AA .}

\figurenum{4b}
\caption{
The dependence of the photospheric temperature, in the neutral case,
on a combination of model parameters for $\lambda=5500\mbox{\AA}.$
The ranges of model parameters are the same as in Figure 2.
There is no photospheric solution when the abscissa is less than about $-1.5$
(representing low densities and high temperatures, the ionized case).}

\figurenum{5a}
\caption{
The (scaled) dilution factor for the ionized case as a function of
photospheric temperature. The behavior at short
wavelengths reflects that of $\chi_{R}/\chi_\nu .$}

\figurenum{5b}
\caption{
The dilution factor as a function of photospheric
temperature for $\alpha=8,$ at three wavelengths.
Both the ionized regime (dashed) and the neutral regime (solid) are
shown. For each wavelength, the upper curve is for $\drpf = 1$ while the
lower curve is for $\drpf = 10 .$}
\end{figure}

\typeout{}
\typeout{YOU MAY OBTAIN THE FIGURES VIA ANONYMOUS FTP TO mensch.stanford.edu.}
\typeout{THE APPROPRIATE FILE IS figs.9408040.ps.uu . THE FILES HAVE BEEN}
\typeout{processed with the uufiles command, and are about 3MB in size.}
\typeout{}
\end{document}